\documentclass{ws-procs9x6}

\begin{document}

\title{Formation of Magnetic Monopoles\\ in Hot Gauge Theories
}

\author{A.~Rajantie}

\address{
DAMTP, CMS, Wilberforce Road, 
\\
University of Cambridge,\\
Cambridge CB3 0WA,
 United Kingdom\\
E-mail: a.k.rajantie@damtp.cam.ac.uk}


\maketitle

\abstracts{
In this talk, I discuss the formation of magnetic monopoles in a phase
transition from the confining SU(2) phase to the Coulomb phase in a
hot Georgi-Glashow model. I argue that monopoles are
formed from
long-wavelength thermal fluctuations, which freeze out after the phase
transition.
}

\vspace*{-9cm}
\begin{flushright}
DAMTP-2003-2
\end{flushright}
\vspace*{7.24cm}

\section{Introduction}

In the non-inflationary Big Bang scenario, 
magnetic monopoles formed at the phase transition of the Grand Unified
Theory (GUT) lead
to a severe cosmological problem, because their energy density would
be so high that the universe would have collapsed under its own weight
a long time ago. This problem is famously solved by inflation,
provided that the temperature of the universe never reaches the GUT
scale after inflation ended. On the other hand, this also gives a
constraint for possible inflationary models.

Most of the work done on the monopole problem deals with the subsequent
annihilations of monopole-antimonopole pairs, but it has been
generally
assumed
that the monopoles were formed by the Kibble mechanism.\cite{Kibble:1976sj} In
that case, causality would give a lower bound for the number density 
of one monopole
per Hubble volume at the time of the GUT transition. However, the
Kibble mechanism implicitly assumes that the symmetry broken in the
transition is a global one, and its validity is gauge field theories,
such as GUTs, is therefore not obvious.

There is also a possibility in many inflationary models that a broken
symmetry is temporarily restored by non-thermal 
fluctuations\cite{Kofman:1995fi}
even if the equilibrium temperature never exceeds the critical
temperature. This would impose even stronger constraints on
inflationary models, but calculating what they really are requires an
understanding of the principles that govern monopole formation more
generally.

In this talk, I will discuss monopole formation in gauge field
theories
and review the basic idea of the scenario presented in
Ref.~\refcite{Rajantie:2002dw}.
This scenario is different from the Kibble mechanism, 
and 
the difference is 
reflected both
in the number density and in the spatial distribution of monopoles.
Therefore, we are not simply using different language to
describe the same physics.

\section{Monopole formation}

For simplicity, let us consider an SU(2) gauge field coupled to an
adjoint Higgs field $\Phi$ at a non-zero temperature. Although realistic GUTs
are much more complicated, this toy model has all the properties that
are important for monopole formation. We imagine that the Higgs
potential has the form
\begin{equation}
V(\Phi)=m^2(t){\rm Tr}\Phi^2 + \lambda {\rm Tr}\Phi^4,
\end{equation}
where $\lambda$ is high enough so that the transition is continuous and
the mass parameter $m^2(t)$ varies with time. We start from
thermal equilibrium at a high enough value of $m^2$ so that the SU(2)
symmetry is unbroken, and then gradually decrease $m^2$ through the
critical point into the broken phase.

If the symmetry were global, one could simply say that in the
broken phase the ground state corresponds to a non-zero $\Phi$, and
because of the SU(2) symmetry, there is a two-sphere of different
vacua. Ideally, $\Phi$ should point into the same direction
everywhere, but that would mean that it is correlated over an infinite
distance, and that cannot be achieved in a finite time. 
When the critical point is approached, the equilibrium correlation
length $\xi$ diverges, but in reality it can only reach some finite value
$\hat\xi$ before the system falls out of equilibrium.\cite{Zurek:1985qw}
Following Kibble,\cite{Kibble:1976sj} we can then argue that
if we consider
two points separated by more than $\hat\xi$,
the choice of the vacuum at these points must be
uncorrelated. This leads to the formation of magnetic monopoles.

In gauge theories, there are extra complications. Because $\Phi$ is
not gauge invariant, neither it nor its correlators are physical observables.
Monopoles do, of course, still 
exist\cite{'tHooft:1974qc,Polyakov:ek} but
we cannot use the above argument to describe their formation.
Furthermore, $\Phi$ cannot be used as an order 
parameter,\cite{Elitzur:im}
either. There is strictly speaking no phase transition
between the ``symmetric'' and the ``broken'' phase in the gauge
theory.\cite{Hart:1996ac}
This means that all correlation lengths remain finite
at the transition point. This is another reason why Kibble's argument
cannot be used in the gauge theory.

It has been suggested that these problems could be avoided by fixing a
gauge so that only a global SU(2) symmetry remains.\cite{} The problem
with this idea is that in a typical fixed gauge, there is no reason
why $\Phi$ should be constant in space even in the broken phase, since
the gauge field can be non-zero. 
On the other hand, one could also fix the unitary
gauge by rotating $\Phi$ everywhere to the same direction, and then a
naive application of the Kibble scenario would lead to the wrong
result that no monopoles are formed. 
Obviously, the gauge field plays
an important role in the problem.

The first step in understanding the dynamics of the phase transition
is understanding the nature of the two phases in equilibrium.
All we actually need to know about the symmetric phase is 
that it is confining and that the longest
correlation lengths are of order $1/g^2T$.
In perturbation theory, there is a massless photon in the broken phase, but
non-perturbative effects make it 
massive,\cite{Polyakov:vu} giving it a magnetic
screening mass $m_B\propto\exp(-m_M/2T)$, where $m_M$ is the monopole
mass. At zero temperature, we would then reach the true Coulomb phase.

One can now see that although there are no diverging correlation
lengths at the transition point, the magnetic screening length grows
faster and faster as we go deeper into the broken phase. If we keep on
decreasing $m^2$ at a constant rate, the screening length
$\xi_B=1/m_B$ would eventually
have to grow faster than the speed of light. This is impossible, and
therefore system must fall out of equilibrium.\cite{Rajantie:2002dw}

The fact that the system falls out of equilibrium does not, of course,
necessarily mean that monopoles are formed. However, the diverging
correlation length here is the magnetic screening length, defined by
\begin{equation}
\langle B_i(\vec{x}) B_j(\vec{y})\rangle \sim
\exp(-m_B|\vec{x}-\vec{y}|).
\end{equation}
Here $B_i$ could, in principle, be any operator that couples to the
photon, but for our purposes it is most convenient to use the 't~Hooft
operator for the magnetic field\cite{'tHooft:1974qc} or its discretized
analogue.\cite{Davis:2000kv} That has the advantage that $B_i$ only has delta
function sources, which can be interpreted as magnetic monopoles.

Using this interpretation, we can define the monopole density $\rho_M$
as
$
\rho_M=\vec{\nabla}\cdot\vec{B},
$
and it follows that the magnetic charge-charge correlator will also
fall exponentially with the same decay rate
\begin{equation}
\langle \rho_M(\vec{x})\rho_M(\vec{y}) \rangle \sim
\exp(-m_B|\vec{x}-\vec{y}|).
\end{equation}
This exponential tail
will become longer as we go deeper into the broken phase and $m_B$
decreases. There will therefore be long-wavelength charge
fluctuations, which will persist and stop the monopole density from
falling to zero.

If we assume that the system stays in equilibrium long enough after
the perturbative transition point, we can actually calculate
the charge-charge correlator\cite{Rajantie:2002dw}
\begin{equation}
\langle\rho_M(\vec{x})\rho_M(\vec{y})\rangle
\approx q_M^2n_M\left(\!
\delta(\vec{x}\!-\!\vec{y})-\frac{m_B^2}{4\pi|\vec{x}\!-\!\vec{y}|}
e^{-m_B|\vec{x}\!-\!\vec{y}|} \!\right)\!.
\label{equ:eqcorrcoord}
\end{equation}
Here $q_M=4\pi/g$ is the magnetic charge and 
\begin{equation}
n_M\approx g^2T/\xi_B^2
\end{equation} 
is the thermal monopole density.

Deeper into the broken phase, the monopoles cost more
energy and the monopole density $n_M$ must therefore decrease. Because
monopoles are stable, this can only take place through annihilations
of monopole-antimonopole pairs. At first, when $\xi_B$ is small
enough, the charge-charge correlation can keep the form
(\ref{equ:eqcorrcoord}) and the
system can stay in equilibrium, because monopoles can travel the
distance $\xi_B$ to be annihilated.

However, sooner or later the distance $\xi_B$ becomes too long for this, 
and the monopoles will instead try to find antimonopoles
closer by. The long-distance part of the correlator
(\ref{equ:eqcorrcoord}) will then
freeze and $\xi_B$ stop to some value $\hat\xi_B$. In the ideal case, all
monopole-antimonopole pairs of size less than  $\hat\xi_B$ are
annihilated, but larger pairs survive.

Basically, this means that the charge distribution at the time of the
freeze-out gets smoothed so that all details at scales shorter than
$\hat\xi_B$ disappear, and the remaining charge distribution is
divided into quantized unit charges, which are the monopoles.

To obtain a rough estimate of the number density of monopoles, 
we can assume that the typical charge inside a
given sphere of radius $\hat\xi_B$ cannot change but gets frozen to the
value it had when the system fell out of equilibrium,
\begin{equation}
Q_M(\hat\xi_B)
= \sqrt{\left\langle \left(\int^{\hat\xi_B} d^3x \rho_M(\vec{x})
\right)^2 \right\rangle}\approx \sqrt{T\hat\xi_B}.
\label{equ:rmscharge}
\end{equation}
This means that in the final state, there will be $Q_M(\hat\xi_B)/q_M$
monopoles inside the sphere, and most strikingly, they would all have
the same sign. If
$Q_M(\hat\xi_B)/q_M$ is large, there will be regions of many monopoles but
no antimonopoles (and, of course, others with many antimonopoles
but no monopoles). In other words, this scenario predicts
positive correlations between monopoles at short distances, whereas
the Kibble mechanism would predict negative
correlations.\cite{Rajantie:2002dw} 

We can also see that because there are roughly $Q_M(\hat\xi_B)/q_M$ monopoles
inside any sphere of volume $\hat\xi_B^3$, the  monopole number
density
will be
\begin{equation}
n_M\approx \frac{Q_M(\hat\xi_B)}{q_M\hat\xi_B^3} 
\approx q_M^{-1}\sqrt{\frac{T}{\hat\xi_B^5}}\approx 
g\sqrt{\frac{T}{\hat\xi_B^5}}.
\label{equ:omapred}
\end{equation}

\section{Conclusions}
In this talk, I have only tried to present the scenario in its
simplicity and not to do any quantitative calculations. Some very
simple estimates can be done easily,\cite{Rajantie:2002dw} but more precise
calculations are going to need a much better knowledge of the
equilibrium and non-equilibrium properties of hot gauge theories. 

Firstly, static equilibrium simulations can be used to measure the
screening length $\xi_B$ or, even better, the charge-charge correlator
as a function of $m^2$. Real-time simulations such as those used to
measure the sphaleron rate can be used to find out when the
freeze-out takes place and determine $\hat\xi_B$. And finally,
simulations of the whole phase transition could be used to test this
scenario. Because the phenomenon is sensitive to non-perturbative
dynamics of a non-Abelian gauge field, simulations like that would be
a great challenge to any methods based on, say,  Hartree or 2PI
formalisms, but should be relatively straightforward to do using the
classical approximation.

\section*{Acknowledgments}
\uppercase{T}his work was supported by 
\uppercase{PPARC}, \uppercase{C}hurchill
\uppercase{C}ollege and the 
\uppercase{ESF COSLAB} programme.

\end{document}